\documentclass[12pt,preprint]{aastex}

\usepackage{graphicx}
\usepackage{txfonts}
\usepackage{longtable}
\usepackage{lscape}
\usepackage{natbib}

\shorttitle{The polarization of the solar Mg~{\sc ii} $h$ and $k$ lines}
\shortauthors{Belluzzi and Trujillo Bueno}

\begin{document}

\title{The polarization of the solar Mg~{\sc ii} $h$ and $k$ lines}

\author{{\sc Luca Belluzzi}\altaffilmark{1,2} {\sc and Javier Trujillo 
Bueno}\altaffilmark{1,2,3}}
\altaffiltext{1}{Instituto de Astrof\'isica de Canarias, E-38205 La Laguna, 
Tenerife, Spain}
\altaffiltext{2}{Departamento de Astrof\'isica, Facultad de F\'isica, 
Universidad de La Laguna, Tenerife, Spain}
\altaffiltext{3}{Consejo Superior de Investigaciones Cient\'ificas, Spain}

\begin{abstract}

Although the $h$ and $k$ lines of Mg~{\sc ii} are expected to be of great 
interest for probing the upper solar chromosphere, relatively little is 
known about their polarization properties which encode the information on 
the magnetic field. 
Here we report the first results of an investigation whose main goal is to 
understand the physical mechanisms that control the scattering polarization 
across these resonance lines and to achieve a realistic radiative transfer 
modeling in the presence of arbitrary magnetic fields. 
We show that the joint action of partial frequency redistribution (PRD) and 
quantum interference between the two excited $J$-levels produces a complex 
fractional linear polarization ($Q/I$) pattern with large polarization 
amplitudes in the blue and red wings, and a negative feature in the spectral 
region between the two lines. 
Another remarkable peculiarity of the $Q/I$ profile is a conspicuous 
antisymmetric signal around the center of the $h$ line, which cannot be 
obtained unless both PRD and $J$-state interference effects are taken 
into account.
In the core of the $k$ line, PRD effects alone produce a triplet peak structure 
in the $Q/I$ profile, whose modeling can be achieved also via the two-level 
atom approximation. 
In addition to the Hanle effect in the core of the $k$ line, we emphasize also 
the diagnostic potential of the circular polarization produced by the Zeeman 
effect in the $h$ and $k$ lines, as well as in other Mg~{\sc ii} lines located 
in their wings.
\end{abstract}


\section{Introduction}
It is through the Zeeman and Hanle effects that the magnetic fields of the 
solar atmosphere leave their fingerprints in the polarization of the emergent 
spectral line radiation (e.g., Casini \& Landi Degl'Innocenti 2007). 
The art of ``measuring" solar magnetic fields thus relies on the development 
of suitable diagnostic tools based on such effects. 
The Zeeman effect produces circular and linear polarization signals whose 
amplitudes are proportional to ${\cal R}$ and ${\cal R}^2$, respectively, with 
$\mathcal R$ the ratio between the Zeeman splitting and the Doppler line-width. 
The Hanle effect is the modification of the linear polarization produced by 
scattering processes in a spectral line, caused by the presence of a magnetic 
field. This line scattering polarization (whose amplitude is proportional to 
the fractional anisotropy of the incident radiation) is sensitive to magnetic 
strengths between approximately 0.2$B_H$ and 5$B_H$, with $B_H$ the 
critical Hanle field for which the Zeeman splitting of the line's level under 
consideration is equal to its natural width.
Therefore, while the Zeeman effect in the far-UV (FUV) and extreme-UV (EUV) 
spectral lines that originate in the outer solar atmosphere is of limited 
practical interest for exploring its magnetism (because ${\cal R}\ll{1}$), the 
Hanle effect is of great diagnostic potential because its magnetic sensitivity 
is independent of the wavelength and Doppler width of the spectral line under 
consideration.

In this Letter we report the first results of a theoretical investigation aimed 
at clarifying the full diagnostic capabilities of the Mg~{\sc ii} $h$ and $k$ 
lines at 2802.7~\AA\ and 2795.5~\AA, respectively, with emphasis on the linear 
polarization pattern produced by scattering processes. 
These strong resonance lines are of interest for probing the 
hot-temperature plasma of the upper solar chromosphere (e.g., Uitenbroek 1997). 
The Mg~{\sc ii} $h$ and $k$ lines result from two transitions between a 
common lower level with angular momentum $J=1/2$ and two upper levels with 
$J=1/2$ and $J=3/2$, respectively. 
As pointed out in previous theoretical investigations 
\citep[e.g.,][]{Aue80,Bel11}, quantum interference between such excited 
$J$-levels is expected to produce observable effects in the wings of these 
lines, such as a cross-over of $Q/I$ about the continuum polarization level 
in the spectral region between the two lines.
This characteristic observational signature of $J$-state interference, 
pointed out by Stenflo (1980) in the Ca~{\sc ii} H and K lines, is 
expected to be more significant for the Mg~{\sc ii} $h$ and $k$ lines because 
their wavelength separation and the ratio of the continuum to line opacity is 
much smaller than in the case of the Ca~{\sc ii} lines 
\citep[e.g., see Fig.~5 of][]{Bel11}.
In fact, using the approximation of coherent scattering {\em in the observer's 
frame}, and simplifying the radiative transfer computations by assuming that the 
line source function and the integrated mean intensity are equal to the Planck 
function, \citet{Aue80} estimated values of $Q/I$ of about 4\% in the positive 
polarization maxima in the far wings, and of about -1.8\% in the negative 
minimum between the two lines, for a line of sight (LOS) with 
$\mu={\rm cos}\,{\theta}=0.1$ (with $\theta$ the heliocentric angle). 

Another key physical ingredient for modeling the Mg~{\sc ii} $h$ and $k$ lines 
is partial frequency redistribution (PRD). 
As shown by radiative transfer calculations carried out using a two-level 
model atom, PRD effects are expected to produce a conspicuous triplet peak 
$Q/I$ signal in the core of the $k$ line (e.g., Sampoorna et al. 2010; see 
also figure~9 in Trujillo Bueno 2011, obtained in collaboration with Drs. M. 
Sampoorna and J. \v{S}t\v{e}p\'an). 
However, since the dispersion wings of these lines are significantly overlapped 
and $J$-state interference cannot be neglected, it is crucial to investigate 
the scattering polarization problem of the Mg~{\sc ii} $h$ and $k$ system 
accounting for the joint action of both PRD and $J$-state interference effects.
As we shall see below, our PRD with $J$-state interference approach to this 
complex theoretical problem is not restricted to the line wings, so that our 
full radiative transfer calculations in model C of Fontenla et al. (1993; 
hereafter FAL-C model) allow us to provide information on the amplitude 
and shape of the $Q/I$ signal throughout the whole profile.
A correct modeling of the scattering polarization in the core of spectral 
lines is particularly important because it is precisely in this spectral line 
region where the Hanle effect operates. 
We also provide estimates on the circular polarization induced by the 
longitudinal Zeeman effect.

\section{Formulation of the problem}
A quantum mechanical derivation of the redistribution matrix for polarized 
radiation, for a two-level atom with unpolarized and infinitely sharp lower 
level, was carried out by \citet{Dom88} and \citet{Bom97a,Bom97b}.
In the atom rest frame (where the frequency and angular dependencies can be
factorized), frequency redistribution is described by the linear combination
of Hummer's $R_{II}$ and $R_{III}$ functions \citep[see][]{Hum62}, which 
describe purely coherent scattering and completely redistributed scattering, 
respectively.
The corresponding branching ratios for the $K$-th multipole component of 
the redistribution matrix are given by:
\begin{eqnarray}
	& & R_{II} : \;
	\alpha = \frac{\Gamma_R + \Gamma_I}{\Gamma_R + \Gamma_I + \Gamma_E} \, , 
	\label{Eq:br_RII} \\
	& & R_{III} : \;
	\beta^{(K)}-\alpha = 
	\frac{\Gamma_R + \Gamma_I}{\Gamma_R + \Gamma_I + D^{(K)}} - 
	\frac{\Gamma_R + \Gamma_I}{\Gamma_R + \Gamma_I + \Gamma_E} \, ,
	\label{Eq:br_RIII}
\end{eqnarray}
where $\Gamma_R$, $\Gamma_I$ and $\Gamma_E$ are the broadening widths of the 
upper level due to radiative decays, superelastic collisions (i.e., collisional 
de-excitation), and elastic collisions, respectively, while $D^{(K)}$ is the 
depolarizing rate due to elastic collisions ($D^{(0)}=0$).
Note that the branching ratios given in Eqs.~(\ref{Eq:br_RII}) and 
(\ref{Eq:br_RIII}) do not include the factor $(1-\epsilon)$, with 
$\epsilon=\Gamma_I/(\Gamma_R+\Gamma_I)$ the photon destruction probability
\citep[cf., Eqs.~(54) and (55) of][]{Bom97b}.
Their sum is equal to 1 for the $K=0$ multipole component 
($\beta^{(0)} \equiv 1$), while in the presence of depolarizing collisions
it is generally smaller than 1 for $K \neq 0$.

As mentioned in Section 1, quantum interference between different $J$ 
levels is expected to produce significant observable effects on the scattering
polarization signal across the Mg~{\sc ii} $h$ and $k$ lines. 
An atomic model accounting for the contribution of $J$-state interference 
\citep[such as the two-term model atom presented in][herefater LL04]{Lan04}
is therefore needed for the investigation of this resonance doublet. 
Unfortunately, a rigorous quantum mechanical derivation of the redistribution 
matrix for such an atomic model, in the presence of elastic and inelastic 
collisions, has not been carried out yet. 
Nevertheless, the limit of coherent scattering {\em in the atom rest frame} is 
a reasonable approximation for modeling the scattering polarization across the 
Mg~{\sc ii} $h$ and $k$ lines.
This can be clearly seen from Fig.~1, where the branching ratio $\alpha$ 
corresponding to the $k$ line (the one corresponding to the $h$ line is 
practically identical) is shown as a function of height in the FAL-C model of 
the solar atmosphere.
As shown in the figure, $\alpha$ is indeed equal to one at the heights where 
the line-core optical depth is unity for an observation at $\mu=0.1$, and it 
is very close to one also at the heights where the optical depth in the 
wavelength interval between the two lines is unity. 

A quantum mechanical derivation of the redistribution matrix for a two-term
atom with unpolarized, infinitely sharp lower levels, in the limit of purely 
coherent scattering in the atom's rest frame has been carried out by 
\citet{Lan97} within the framework of the metalevels theory.
This redistribution matrix is suitable for investigating the scattering 
polarization across the Mg~{\sc ii} $h$ and $k$ lines, since the lower level 
of these lines is the ground level (it thus satisfies the hypothesis of an 
infinitely sharp lower level), and it has $J_{\ell}=1/2$.
Such redistribution matrix was obtained by neglecting collisions, and is valid 
in the atom's rest frame. 
Following a derivation analogous to the one presented in \citet{Hum62}, we 
have calculated the corresponding expression in the observer's frame, taking 
Doppler redistribution into account. An alternative derivation has
been recently carried out by \citet{Smi11} starting from the Kramers-Heisenberg 
scattering formula.

We consider a two-term model atom for Mg~{\sc ii}, the lower term being 
composed of the ground level ($^2S_{1/2}$), the upper term of the upper 
levels of the $h$ and $k$ lines ($^2P_{1/2}$ and $^2P_{3/2}$, respectively). 
Although a rigorous derivation of the collisional rates in a two-term atom has 
not been carried out yet, we included the effect of (isotropic) inelastic and 
superelastic collisions in the redistribution matrix, under the following 
assumptions. 
We first assumed that the relaxation rate due to superelastic collisions is 
the same for the two upper $J$-levels (see Section 7.13 of LL04 for the 
explicit expressions of such rate).
We then assumed that such collisional relaxation rate is also the one for the 
interference between the two upper $J$-levels.
Accordingly with the previous approximation, we calculated the inelastic 
collision transfer rates from the common lower level to the two upper levels.
Since the lower level is unpolarized, these latter collisions only affect 
the populations of the upper $J$-levels, and the corresponding rates are
identical to the case of a multi-level atom (see Sect. 7.13 of LL04). 
Collisional rates between pairs of $J$-levels pertaining to the same term
have been neglected, as well as elastic collisions.
Under these hypotheses, following the convention according to which primed 
quantities refer to the incident photon, while unprimed quantities to the 
scattered photon, indicating with $\nu$ and $\vec{\Omega}$ the photon's 
frequency (in the observer's frame) and propagation direction, we obtained
\begin{eqnarray}
	R_{ij}(\nu^{\prime},\vec{\Omega}^{\prime}; \nu, \vec{\Omega}) 
	& = & 3 \, \frac{2L_u+1}{2S+1} \sum_{K} \,
	\sum_{J^{}_{\!u} J^{\prime}_{\!u}} \, 
	\sum_{J^{}_{\!\ell} J^{\prime}_{\!\ell}} \,
	(-1)^{J^{}_{\!\ell} - J^{\prime}_{\!\ell}} \nonumber \\
	& & \times \, (2J^{}_{\!u} +1) \, (2J^{\prime}_{\!u} + 1) \, 
	(2J^{}_{\!\ell} +1) \, (2J^{\prime}_{\!\ell} +1) \nonumber \\
	& & \times \, \bigg\{ 
	\begin{array}{c c c}
		L_u & L_{\ell} & 1 \\
		J^{}_{\!\ell} & J^{}_{\!u} & S
	\end{array}
	\bigg\}
	\bigg\{
	\begin{array}{c c c}
		L_u & L_{\ell} & 1 \\
		J^{}_{\!\ell} & J^{\prime}_{\!u} & S
	\end{array}
	\bigg\}
	\bigg\{
	\begin{array}{c c c}
		L_u & L_{\ell} & 1 \\
		J^{\prime}_{\!\ell} & J^{}_{\!u} & S
	\end{array}
	\bigg\}
	\bigg\{
	\begin{array}{c c c}
		L_u & L_{\ell} & 1 \\
		J^{\prime}_{\!\ell} & J^{\prime}_{\!u} & S
	\end{array}
	\bigg\} \nonumber \\
	& & \times \, 
	\bigg\{
	\begin{array}{c c c}
		K & J^{\prime}_{\!u} & J^{}_{\!u} \\
		J^{}_{\!\ell} & 1 & 1
	\end{array}
	\bigg\}
	\bigg\{
	\begin{array}{c c c}
		K & J^{\prime}_{\!u} & J^{}_{\!u} \\
		J^{\prime}_{\!\ell} & 1 & 1
	\end{array}
	\bigg\} 
	\left[ P^{\, (K)}(\vec{\Omega}^{\prime}, \vec{\Omega}) \right]_{ij}
	\nonumber \\
	& & \times \frac{1}{\pi \, \Delta \nu^2_{\!D} \sin \theta} 
	\, {\rm exp} \left[ -\frac{(\nu^{\prime}-\nu-
	\nu_{J^{\prime}_{\!\ell} J^{}_{\ell} })^2}
	{4 \, \Delta \nu^{2}_{\!D} \sin^2(\theta/2)} \right] \nonumber \\
	& & \times \frac{1}
	{1 + \epsilon^{\prime} 
	+ {\rm i} \frac{\nu_{J^{\prime}_{\!u} J^{}_{\!u}}}{2 \Gamma} } 
	\,\frac{1}{2} \,\bigg[ W \left( \frac{a}{\cos(\theta/2)},
	\frac{\varv^{}_{J^{}_{\!u} J^{}_{\!\ell}} +
	\varv^{\prime}_{J^{}_{\!u} J^{\prime}_{\!\ell}}}
	{2\cos(\theta/2)} \right) \nonumber \\
	& & + W \left( \frac{a}{\cos(\theta/2)},
	\frac{ \varv^{}_{J^{\prime}_{\!u} J^{}_{\!\ell}} +
	\varv^{\prime}_{J^{\prime}_{\!u} J^{\prime}_{\!\ell}}}
	{2\cos(\theta/2)} \right)^{\ast} \bigg] \, ,
	\label{Eq:redistribution}
\end{eqnarray}
where $\left[ P^{(K)} (\vec{\Omega}^{\prime}, \vec{\Omega}) \right]_{ij}$ is 
the $K$-th multipole component of the scattering phase matrix ($i,j=0,1,2,3$), 
$\theta$ is the scattering angle, $\Delta \nu_D$ is the Doppler width (assumed 
to be the same for the two lines), and 
$\epsilon^{\prime}=\Gamma_R/\Gamma_I=A_{u \ell}/C_{u \ell}$, with $A_{u \ell}$ 
the Einstein coefficient for spontaneous emission from the upper to the lower 
term, and $C_{u \ell}$ the superelastic collision relaxation rate.
The function $W$ is defined as 
\begin{equation}
	W(a,\varv)=H(a,\varv) + {\rm i}L(a,\varv) \, ,
\end{equation}
where $H$ is the Voigt function, and $L$ the associated dispersion profile.
The reduced frequencies $\varv_{J_a J_b}$ and $\varv_{J_a J_b}^{\prime}$ are
given by
\begin{equation}
	\varv_{J_a J_b} = \frac{(\nu_{J_a J_b}-\nu)}{\Delta \nu_D} \;\; , 
	\;\;\;\;\;\;\;
	\varv_{J_a J_b}^{\prime} = \frac{(\nu_{J_a J_b}-\nu^{\prime})}
	{\Delta \nu_D} \;\; ,
\end{equation}
with $\nu_{J_a J_b}$ the Bohr frequency between the levels $J_a$ and $J_b$. 
Although elastic collisions have been neglected (consistently with the 
assumption of purely coherent scattering in the atom rest frame), we took 
into account their broadening effect in the evaluation of the broadening 
constant $\Gamma$ appearing in Eq.~(\ref{Eq:redistribution}), and of the 
damping parameter $a=\Gamma/\Delta \nu_{D}$.

\section{The scattering polarization pattern across the Mg~{\sc ii} $h$ and 
$k$ lines}
In order to estimate, in the absence of magnetic fields, the amplitude and 
shape of the scattering polarization pattern across the Mg~{\sc ii} doublet, 
we have developed a non-LTE radiative transfer code based on the angle averaged 
expression \citep[e.g.,][]{Ree82} of the redistribution matrix described in 
the previous section, including the contribution of an unpolarized continuum. 

The numerical method of solution will be explained in detail in a forthcoming 
publication; it is based on a direct generalization to the PRD case of the 
Jacobian iterative scheme presented in Trujillo Bueno \& Manso Sainz (1999).
The initialization of the iterative calculation was done using the 
self-consistent solution of the corresponding unpolarized problem, which we 
have obtained by applying Uitenbroek's (2001) radiative transfer code. 
This code has also been used to compute the inelastic and elastic collisional 
rates, as well as the continuum total opacity (including the UV line haze 
contribution) and emissivity.  

In Figure 2 we show the results of the following PRD calculations of the 
$Q/I$ profile of the emergent radiation for a line of sight with $\mu=0.1$:
(1) the full two-term atom solution obtained by taking into account the impact 
of interference between the two excited $J$-levels, (2) the two-term atom 
solution obtained by neglecting $J$-state interference effects, and (3) the 
solution for the Mg~{\sc ii} $k$-line alone assuming a two-level atom with 
$J_l=1/2$ and $J_u=3/2$. 

The left panel of Fig.~2 shows the overall structure of the $Q/I$ profile. 
Consider first the two-term atom solutions obtained by including (solid line)
and neglecting (dashed line) $J$-state interference. 
Clearly, the impact of $J$-state interference is very important, since it leads 
to much larger polarization amplitudes in the wings (about 10\% at around 
${\pm}15$~\AA\ from the $h$ and $k$ line centers, respectively) and to a 
significant negative polarization feature between the $h$ and $k$ lines with a 
maximum amplitude of about -2\%.
The two-level atom solution for the $k$ line (dotted line) produces wing 
polarization signals which are also significantly smaller than those 
corresponding to the full solution.

The right panel of Fig.~2 shows the details of the $Q/I$ profile in the core 
regions of the Mg~{\sc ii} $h$ and $k$ lines. 
The three solutions perfectly agree in the core of the $k$ line, with a 
line-center amplitude of about 2\%. 
However, it can be observed that the solution obtained by neglecting $J$-state 
interference (dashed line) does not show either the small asymmetry in the two 
negative $Q/I$ peaks around the $k$-line core, or the clear antisymmetric $Q/I$ 
feature that the combined action of PRD and $J$-state interference effects 
produces around the center of the $h$ line (whose polarizability is zero).

Although the hypothesis of purely coherent scattering in the atom's rest frame 
is a good approximation in the core of the lines and between them, 
frequency redistribution effects due to elastic collisions might not be 
completely negligible going towards the wings of the lines (see Fig.~1).
In order to have a qualitative idea of the impact of this physical ingredient, 
we carried out a calculation including the contribution of the $R_{III}$ 
redistribution function, under the limit of complete frequency redistribution 
(CRD) in the observer's frame.
We derived the expression of $R_{III}$ from the CRD theory presented in LL04, 
including the effect of inelastic and superelastic collisions as in the case 
of $R_{II}$.
We used the branching ratios $\alpha$ and $(\beta^{(K)} - \alpha$) defined in 
Eqs.~(1) and (2) (strictly valid in the case of a two-level atom)\footnote{Note 
that the factor $(1-\epsilon)$ is included in our $R_{II}$ and $R_{III}$
redistribution functions.}. 
We assumed $D^{(2)}=0$ (i.e. we neglected the depolarizing effect of elastic 
collisions).
As expected, the contribution of $R_{III}$ is negligible in the core of the 
lines, and very small between them. However, as shown in Fig.~3, it modifies 
in an appreciable way the amplitude of the $Q/I$ profile in the wings 
of the lines.
Note that the solution corresponding to the approximation of purely coherent 
scattering in the observer's frame is also shown in Fig. 3 (see the 
filled circles).

\section{Concluding comments}
As shown in this Letter, the physics that outside sunspots controls the 
scattering polarization of the Mg {\sc ii} $h$ and $k$ lines is the joint 
action of PRD and $J$-state interference effects, which produces a complex 
$Q/I$ pattern with sizable polarization maxima in the wings, and a negative 
polarization feature at wavelengths between the two line centers (see Fig.~2). 
These $Q/I$ wing features can also be understood using the approximation of 
coherent scattering in the observer's frame (see the filled circles in 
Fig.~3)\footnote{It is interesting to note that the observations performed by  
Henze \& Stenflo (1987) could not confirm the negative polarization 
minimum between the $h$ and $k$ lines. 
Although the statistical significance of such pioneering observations is very 
marginal, in a forthcoming investigation we will include the interaction with 
the continuum polarization processes in order to study its possible impact on 
the polarization amplitudes across these lines.}. 
Another remarkable feature of the $Q/I$ profile we have obtained is a clear 
antisymmetric $Q/I$ signal around the center of the $h$ line (see Fig.~2), 
which can only be found when taking into account the joint action of PRD and 
$J$-state interference effects. 
In the core of the $k$ line, PRD effects alone are responsible of the triplet 
peak structure of the $Q/I$ profile (see Fig.~2).

The reported linear polarization is sensitive to a magnetic field through  
the Hanle effect, which operates only in the core of the $k$ line. Here
magnetic fields weaker than 100~G are expected to produce significant changes 
in the line-center polarization amplitude.  
As shown in Fig.~2, it is a good new that the $Q/I$ profile around the core of 
the Mg~{\sc ii} $k$-line can be well modeled through the two-level atom 
approximation, since this will facilitate the development of Hanle-effect 
diagnostic tools. 
In addition to the Hanle effect in the $k$-line, complementary information on 
the magnetic field of the solar chromosphere could be obtained by measuring the 
circular polarization produced by the Zeeman effect in all the 
Mg~{\sc ii} lines that are located between 2790~\AA\ and 2805~\AA\ (see caption 
of Fig. 1). 
The Stokes $V$ signals of all these lines should be measurable. 
For example, a volume filling magnetic field of 100 G inclined by $45^{\circ}$ 
with respect to the LOS produces $V/I$ amplitudes of about 1\% in all these 
Mg~{\sc ii} lines.

Spectroscopy of the Mg~{\sc ii} $h$ and $k$ lines with novel space telescopes 
like the Interface Region Imaging Spectrograph \citep[IRIS; see][]{Tit12} will 
provide precious information on temperatures, flows and waves. 
However, the magnetic field information is encoded in the spectral line 
polarization, whose measurement with high spatial and/or temporal resolution 
requires the development of larger aperture telescopes such as SOLAR-C 
\citep[see][]{Shi11}. 
The results shown in this Letter for the $h$ and $k$ lines of Mg~{\sc ii}, and 
in \citet{JTB11a,JTB12} for the Ly$\alpha$ lines of H~{\sc i} and He~{\sc ii}, 
strongly encourage the development of FUV and EUV spectro-polarimeters for the 
new generation of solar space telescopes. 
Additional results obtained in different atmospheric models, along with the 
details of the methods used in this investigation, will be presented in a 
forthcoming paper.

\acknowledgements
Financial support by the Spanish Ministry of Economy and Competitiveness 
through projects AYA2010-18029 (Solar Magnetism and Astrophysical 
Spectropolarimetry) and CONSOLIDER INGENIO CSD2009-00038 (Molecular 
Astrophysics: The Herschel and Alma Era) is gratefully acknowledged.

\newpage

\begin{figure}[t]
\centering
\includegraphics[width=\textwidth]{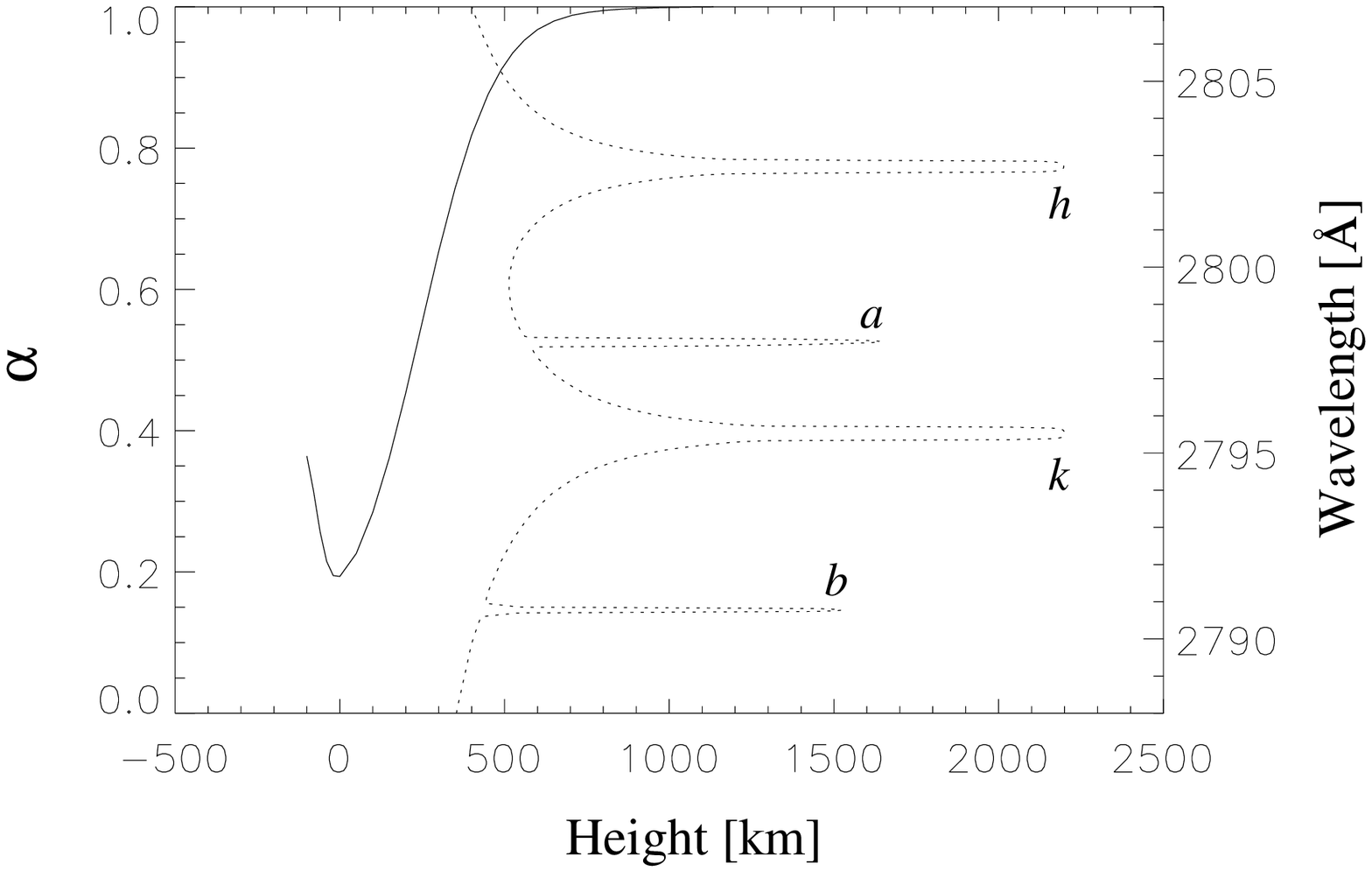}  
\caption{The solid line shows the variation with height in the FAL-C model 
atmosphere of the branching ratio $\alpha$ of Eq.~(1). The dotted line gives 
the atmospheric height where, for each wavelength, the ensuing optical depth 
is unity along a LOS with $\mu=0.1$, assuming that the only blends in the wings 
of the $h$ and $k$ lines are those produced by the following Mg~{\sc ii} lines: 
($a$) two blended lines at 2797.930~\AA\ and 2797.998~\AA\ (resulting from the 
$^2{\rm P}_{3/2}\,-\,^2{\rm D}_{3/2}$ and $^2{\rm P}_{3/2}\,-\,^2{\rm D}_{5/2}$ 
transitions), and ($b$) one at 2790.777~\AA\ (caused by the 
$^2{\rm P}_{1/2}\,-\,^2{\rm D}_{3/2}$ transition). 
Note that these absorption lines of Mg~{\sc ii} originate a few hundred 
kilometers below the Mg~{\sc ii} $h$ and $k$ lines, but they are still 
chromospheric lines.}
\label{fig:figure-1}
\end{figure}

\clearpage 

\begin{figure}[t]
\centering
\includegraphics[width=12cm]{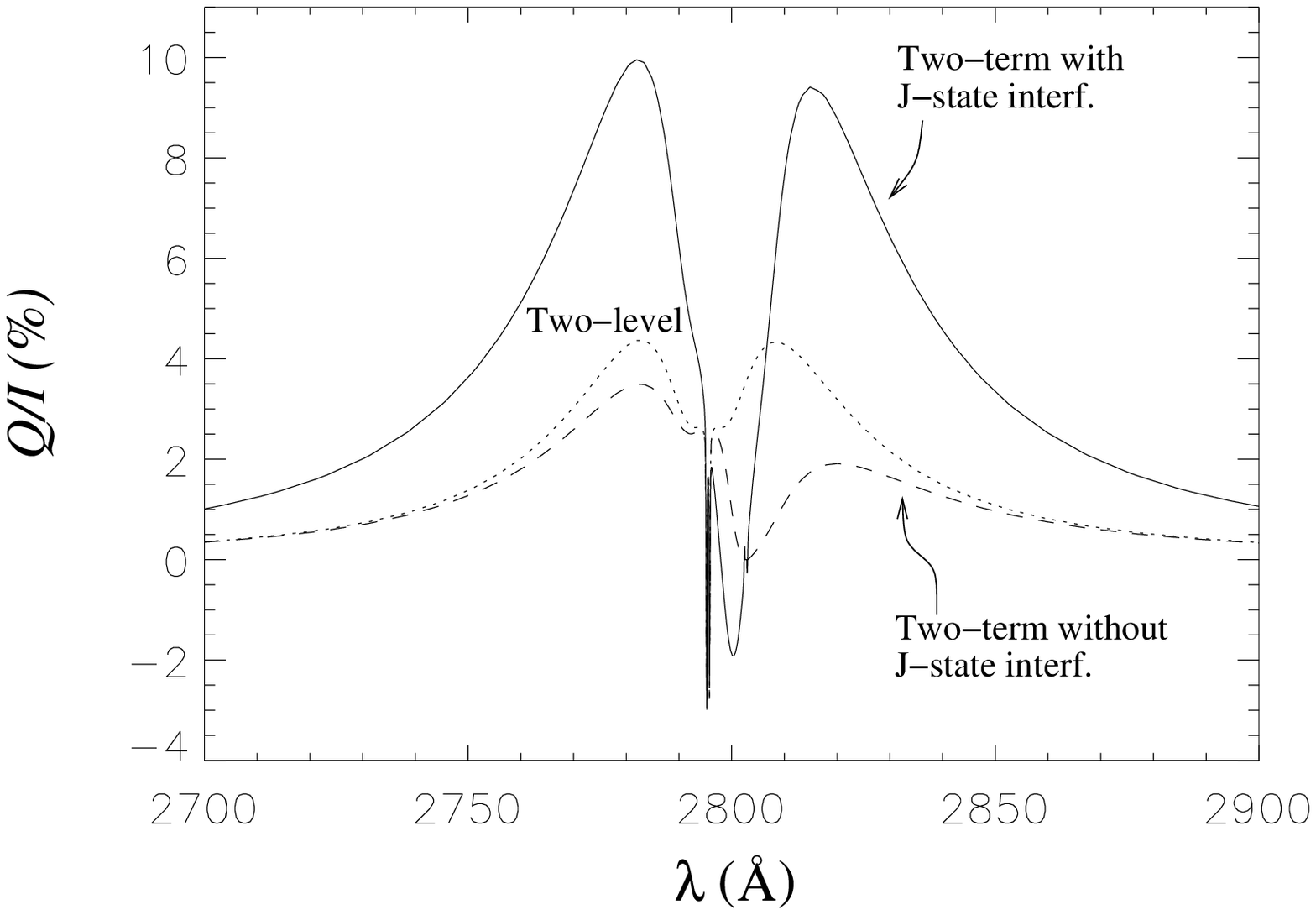}  
\includegraphics[width=12cm]{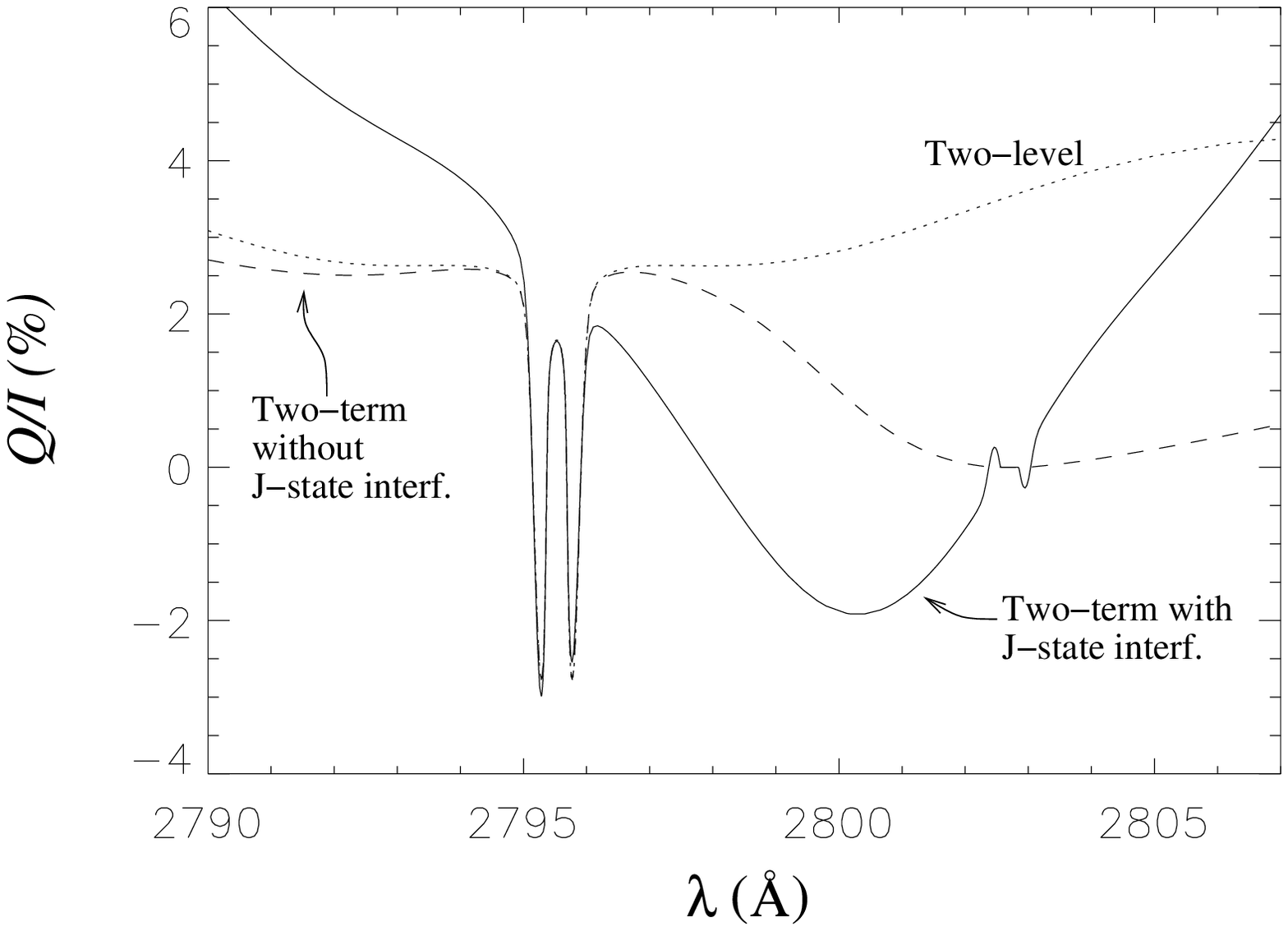}
\caption[]{The $Q/I$ profile across the Mg~{\sc ii} $h$ and $k$ lines, 
calculated in the FAL-C model atmosphere for a LOS with $\mu=0.1$. 
Solid line: two-term atom PRD solution with $J$-state interference. 
Dashed line: two-term atom PRD solution neglecting $J$-state interference. 
Dotted line: two-level atom PRD solution for the $k$-line transition. 
The reference direction for positive $Q$ is the parallel to the nearest limb. 
The right panel shows in more detail the line core regions.}
\label{fig:figure-2}
\end{figure}

\clearpage

\begin{figure}[t]
\centering
\includegraphics[width=12cm]{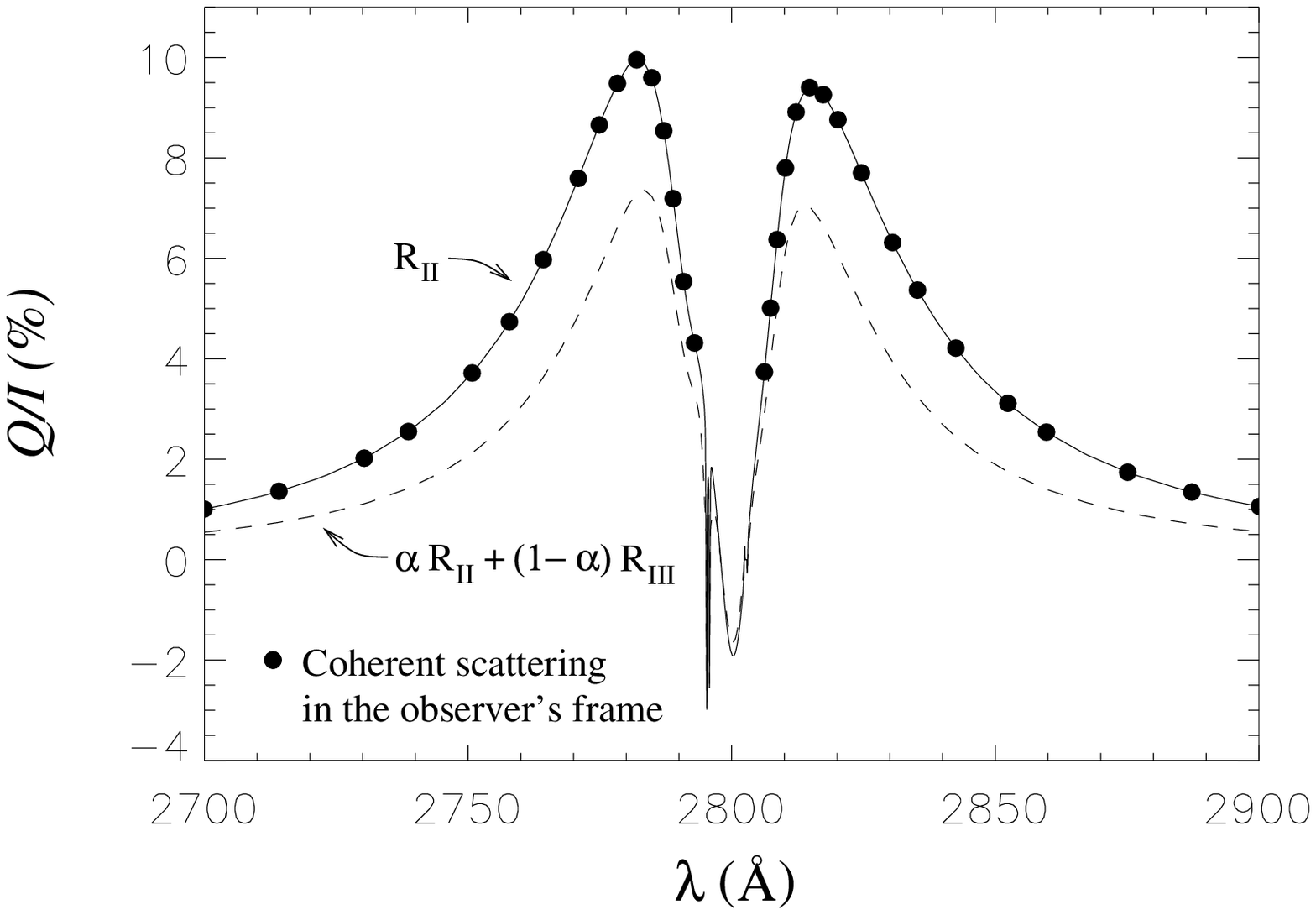}  
\includegraphics[width=12cm]{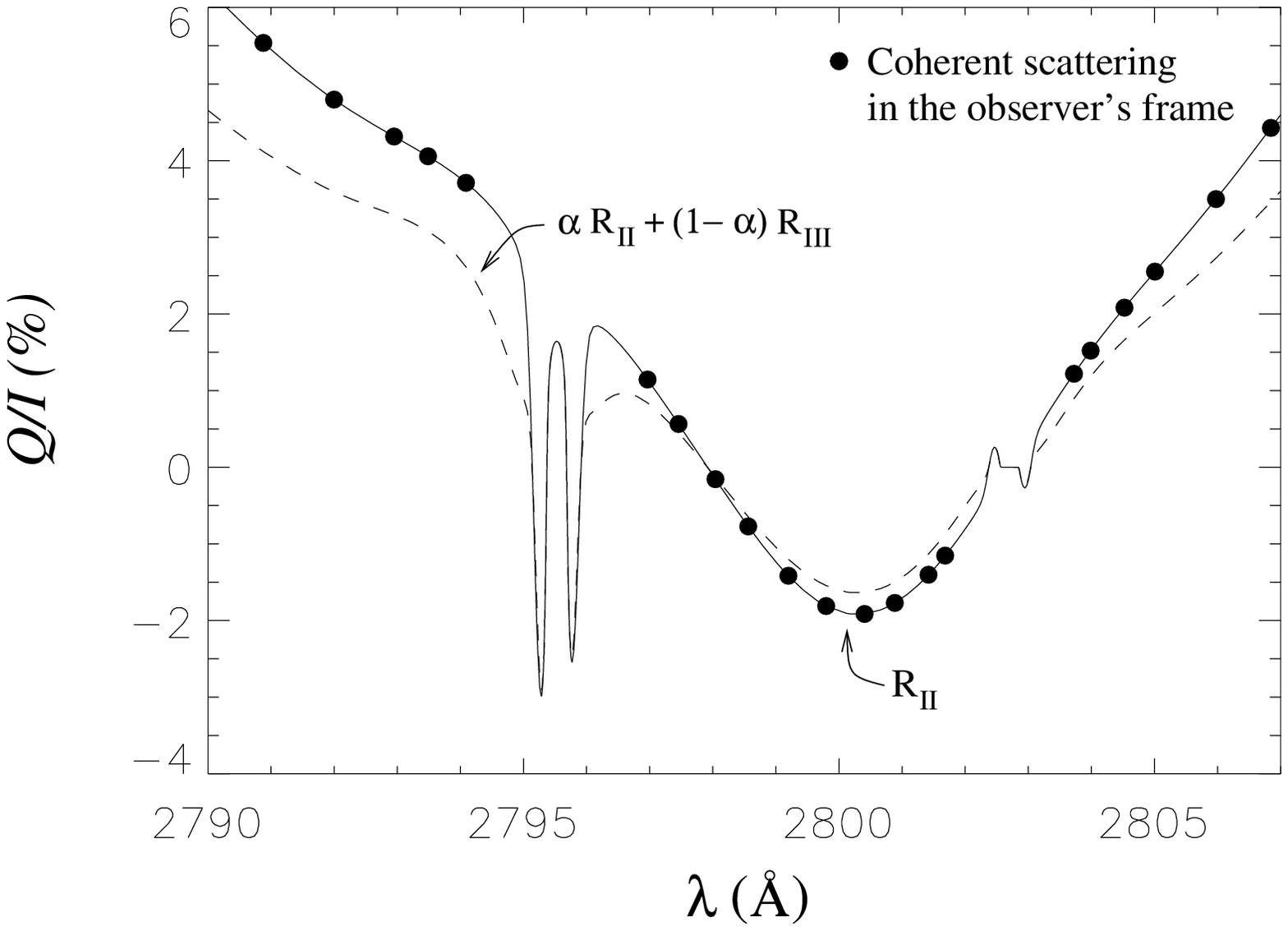}
\caption[]{Comparison between the $Q/I$ profiles obtained by considering only 
the $R_{II}$ redistribution function (solid line, same as in Fig.~2), and a 
linear combination of $R_{II}$ and $R_{III}$ (dashed line). 
The filled circles give the $Q/I$ values obtained under the 
approximation of coherent scattering in the observer's frame, after excluding 
the core regions where such an approximation to the full $R_{II}$ solution does 
not hold.}
\label{fig:figure-3}
\end{figure} 

\end{document}